\documentclass[11pt,titlepage]{article}

\usepackage{amsmath}

\usepackage[round,numbers,sort&compress]{natbib} 
\usepackage{epsfig}
\usepackage{graphicx}
\usepackage{pifont}
\usepackage{amssymb}
\usepackage{latexsym}
\usepackage{verbatim}
\usepackage{array}
\usepackage{xspace}
\usepackage{colordvi}
\usepackage{wasysym}
\usepackage{marvosym}
\usepackage{setspace}
\usepackage{stevedef}
\usepackage{color}
\usepackage{hhline}
\usepackage{rotating}
\usepackage{subfigure}

\newlength{\saveparindent}
\newcounter{saveeqn}%
\newcommand{\alpheqn}{\setcounter{saveeqn}{\value{equation}}%
\stepcounter{saveeqn}\setcounter{equation}{0}%
\renewcommand{\theequation}{\mbox{\arabic{saveeqn}\alph{equation}}}}%
\newcommand{\reseteqn}{\setcounter{equation}{\value{saveeqn}}%
\renewcommand{\theequation}{\arabic{equation}}}%

\begin{document}

\title{Structural alignment using the generalized Euclidean distance between conformations}



\author{Ali~R.~Mohazab \\
	Department of Physics and Astronomy, \\
	University of British Columbia, Vancouver, BC, Canada 
	\and Steven~S.~Plotkin\footnote{Corresponding author.  Address: 
          Department of Physics and Astronomy,
	   University of British Columbia,
	   6224 Agricultural Road, 
	   Vancouver, BC V6T1Z1, Canada,
	   Tel.:~(604)822-8813, Fax:~(604)822-5324, email:~steve@phas.ubc.ca} \\
	Department of Physics and Astronomy, \\
	University of British Columbia, Vancouver, BC, Canada}

\date{}

\pagestyle{myheadings}
\markright{Minimal distance structural alignments} %


\pagenumbering{arabic}
\setcounter{page}{1}


\maketitle
\normalsize

\begin{abstract}
  The usual Euclidean distance may be generalized to extended objects
  such as polymers or membranes. Here  this
  distance is used for the first time as a cost function to align
  structures. We examined the alignment of extended strands to
  idealized beta-hairpins of various sizes using several cost
  functions, including RMSD, MRSD, and the minimal distance. 
  We find that using minimal distance as a cost function typically
  results in an aligned structure which is globally different 
  than that given by an RMSD-based alignment.

\end{abstract}


\section*{Introduction}

In a series of experiments starting in the late 1950s and culminating
in a 1961 paper in the {\it Proceedings of the National Academy of
  Sciences}~\cite{Anfinsen61}, C. B. Anfinsen showed that a protein
such as bovine pancreatic ribonuclease would, under oxidizing
conditions, undergo slow but spontaneous reshuffling of disulfide
bonds from a state with initially random cross-linked pairs, to a
state with correct disulfide pairing and full enzymatic activity. The
spontaneous formation of correct disulfide pairs indicated that the
amino acid sequence itself was guiding the process towards more
thermodynamically favorable configurations, and the so-called
thermodynamic hypothesis in protein folding was born.

This discovery underpinned the formalism developed decades later to
understand protein folding as a configurational diffusion process on
an energy landscape that through molecular evolution had the overall
topography of a rugged
funnel~\cite{WolynesPG92:spinb,Bryngelson95,Wolynes95,Onuchic97,FershtAR99:book,DobsonCM03,PlotkinSS02:quartrev1,PlotkinSS02:quartrev2}. 
The initial
random crosslinkings and subsequent slow exchange of disulfide bonds
observed by Anfinsen and colleagues argued against a mechanistic
pathway picture, but there was nevertheless a lag phase before the
energy landscape picture eventually took hold. 

Though important as a conceptual tool, real predictive power was
brought to bear by quantifying the funnel notion to generate free
energy surfaces as a function of a progress coordinate that
measured the degree to which a protein was
folded~\cite{BryngelsonJD89,Sali94a,PlotkinSS97}. Soon thereafter, 
questions arose regarding what coordinate(s) best represented folding
progress, or whether one could even find a simple geometric coordinate
that would represent kinetically how folded a protein was. The kinetic
proximity of a given configuration was quantified unambiguously as the
probability a protein would fold first before unfolding, given that it
was initially in that given configuration~\cite{DuR98:jcp}. This
idea had earlier analogues  in the Brownian analysis of
escape and recombination probabilities of an ionized
electron~\cite{OnsagerL38}. 

\subsection*{Order parameters in protein folding}

The study of various order parameters that might best represent
progress in the folding reaction have generated much
interest~\cite{Garcia:92,ChanHS94,PlotkinSS98,DuR98:jcp,HummerG01,BaumketnerA04,MaA05,BestRB05,DasP06,ChoSS06,DokholyanNV02,ChavezLL04,WangJ06,SegaM07,BeckDAC07,KrivovSV08},
with questions focusing on what parameter(s) or principle
component-like motions might best correlate with splitting probability
or probability of folding before unfolding.

On the other hand, analyses using intuitive geometric order parameters
have been developed to understand folding and are now commonly
used. These include the fraction of native
contacts $Q$~\cite{NymeyerH00:pnas,PlotkinSS02:Tjcp,WindAF02,CheungMS02,KaranicolasJ02,BestRB05,ChoSS06},
which can be locally or globally defined, 
root mean square distance or deviation (RMSD) between
structures~\cite{PandeGruebeleNature2002,SimmerlingC02,GarciaAE03,HummerG00},
structural overlap parameter
$\chi$~\cite{VeitshansT96,BaumketnerA02,CheungMS06}, Debye-Waller
factors~\cite{PortmanJJ01,PortmanJJ01:jcp}, or fraction of correct Dihedral
angles~\cite{KaranicolasJ02}. 

To find a simple geometrical
order parameter that quantifies progress to
the folded structure poses several challenges.
These include an accurate account of the effects of polymer
non-crossing~\cite{MohazabAR08:bj}, energetic and entropic 
heterogeneity in native driving
forces~\cite{PlotkinSS00:pnas,PlotkinSS02:Tjcp,LindbergM02}, as well
as non-native  
frustration and trapping~\cite{PlotkinSS01:prot,PlotkinSS03,ClementiC04}.
Fortunately it has been borne out experimentally that wild type proteins
are sufficiently minimally frustrated that non-native interactions do
not play a strong role in either folding rate or mechanism, and native
structure based models 
for folding rates and mechanisms have enjoyed considerable
success~\cite{ShoemakerBA97,ShoemakerWang99,MunozV99,Galzitskaya99:PNAS,AlmE99,Baker2000:Nature}.

In condensed matter systems, useful order parameters have historically
had intuitive geometrical interpretations. Their definition did not
require the knowledge of a particular Hamiltonian (although their
temperature-dependence and time-evolution were affected by the
energy function in the system). In chemical reactions, the distance
between constituents in reactant and product has played a ubiquitous
role in the construction of potential energy
surfaces~\cite{LevineRD87}. Moreover from the point of view of
stochastic escape and recombination, the distance perfectly correlates
with the commitment probability for a freely diffusing particle
between two absorbing boundaries.  

\subsection*{Distance as an order parameter}

The distance is easy to define for a point particle, which we imagine
to travel between two locations $A$ at ${\bf r}_A$ and $B$ at ${\bf
  r}_B$. It is the variational minimum of the functional:
\begin{equation}
  \label{eq:pointDistance}
   \int_{{\bf r}_A}^{{\bf r}_B} ds = \int_0^T dt
  \sqrt{\dot{\bf r}^2} 
\end{equation}
where $\dot{\bf r} \equiv d{\bf r}/dt$, and the initial and final conditions, or
equivalently boundary conditions, are ${\bf r}^\ast(0) = {\bf r}_A$ and
${\bf r}^\ast(T) = {\bf r}_B$.

However until recently~\cite{PlotkinSS07,MohazabAR08,MohazabAR08:bj}
the distance had not been formulated for higher 
dimensional objects such as pairs of polymer configurations, despite
close parallels in string theory~\cite{ZwiebachB04}. 

In this paper after briefly reviewing two common reaction
coordinates, Q and RMSD, and the two newer ones introduced and
explored in 
~\cite{PlotkinSS07,MohazabAR08,MohazabAR08:bj}, $\cal D$ and Mean root
squared distance (MRSD), we will further explore structural alignments
based on $\cal D$ for idealized hairpins.

\subsection*{Some problems with commonly used reaction coordinates}

Many reaction coordinates have been used to describe the
folding process, while still being flawed in principle. These
characterizations have been largely successful because the majority of
conformations during folding are well characterized by changes in
these parameters: Proteins undergo some collapse concurrently with
folding, lower their internal energy, and adopt structures
geometrically similar to the native structure. 

Nevertheless it is easy to point to simple examples of conformational
transitions for which the adoption of native structure does not
correlate with the change in commonly used order parameters. While
these conformational pairs may not be wholly representative of the
total folding process, they point to situations where folding to a
given structure would
not be well-characterized by commonly used order parameters.

Figure~\ref{fig:QFlaw1} shows two structures $A$ and $B$ with different measures
of structural similarity to a ``native'' hairpin fragment $N$. These
structures have different measures of proximity depending on the
coordinate used to characterize them. 
If we use the fraction of native contacts $Q$ to describe native
proximity~\footnote{$Q_{\mbox{\tiny{AN}}} \equiv \left( \sum_{i<j}
  \Delta_{ij}^{\mbox{\tiny{A}}} \Delta_{ij}^{\mbox{\tiny{N}}}\right)/
  \left(\sum_{i<j}\Delta_{ij}^{\mbox{\tiny{N}}}\right)$ counts
  pairs of residues with some cutoff distance in both structure $A$
  and structure $N$. This result is then normalized by the number of
  contacts in the native structure.}, structure $A$ has a $Q$ of $Q_{\mbox{\tiny{A}}} = 1/3$
while $Q_{\mbox{\tiny{B}}} = 0$, so by this measure it is more
native. If we use the root mean square deviation RMSD~\footnote{$RMSD
  \equiv \sqrt{N^{-1}\sum_{i = 1}^N ({\bf r}_{A_i} - {\bf
      r}_{B_i})^2}$ is a least-squares measure of similarity between
  structures $A$ and $B$. Typically this quantity is minimized given
  two structures and so can be thought of as a ``least squares
  fit''. The sum may be over all atoms, or simply all residues in
  coarse-grained models.}, structure $B$ is more native-like than $A$. Moreover,
structure $B$ would have a higher probability of folding before
unfolding than $A$, i.e. it has a larger value of
$p_{\mbox{\tiny{FOLD}}}$~\cite{DuR98:jcp}, and so is closer
kinematically to the native structure. The longer the hairpin, the
more likely a slightly expanded structure is to fold, so the
discrepancy between $Q$ and $RMSD$ for these pairs of structures becomes
even larger. 

\begin{figure} [hp]
  \centering
  \includegraphics[width=6cm]{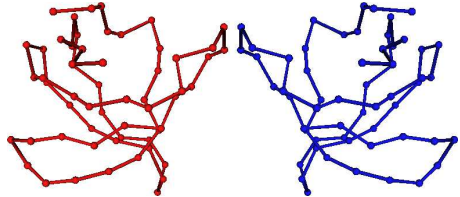}
  \caption{Native structure of SH3 (right) and its mirror image.
    Although dissimilar by RMSD, biologically
    nonfunctional, and disallowed by true dihedral potentials, this
    structure has a $Q=1$, because native contacts remain intact after mirroring
    transformations.}  
  \label{fig:QFlaw2}
\end{figure}

In contrast to $RMSD$, $Q$ also does not distinguish between
chiralities. Typically the energy function forbids opposite
chiralities, however if the appropriate chirality is not enforced in
the backbone dihedral potentials, mirror-image structures as in
figure~\ref{fig:QFlaw2}  will be
allowable, and are indistinguishable according to
$Q$~\cite{MohazabAR08}.

\begin{figure}[hp]
  \centering
  \includegraphics[width=7cm]{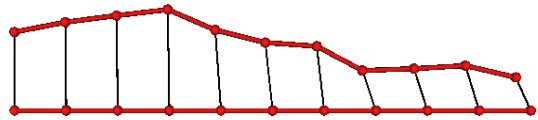}
  \caption{The MRSD is the average length of the black like segments
    between corresponding residues of the initial and final
    configuration.}
  \label{fig:MRSDMeaning}
\end{figure}

While the RMSD is often characterized as a ``distance'' between
structures, it is not equivalent or even proportional to the 
sum of the straight-line distances between the atoms or residues in
the two structures (figure~\ref{fig:MRSDMeaning}). 
This quantity is in fact given by the 
mean root squared distance (MRSD), defined 
for two structures $A$ and $B$ as:
\begin{equation}
  \label{eq:MRSDDef}
  {1 \over N}\sum_{n = 1}^N |{\bf r}_{A_n} - {\bf r}_{B_n}| = \
  {1 \over N}\sum_{n = 1}^N \sqrt{({\bf r}_{A_n} - {\bf r}_{B_n})^2}
\end{equation}
The $RMSD$ between two structures is always greater than or equal to
the $MRSD$ between the same structures, with $MRSD =
RMSD$ in only the most trivial cases~\cite{MohazabAR08}. The $RMSD$ is
also less robust to large fluctuations of select residues in
structural pairs~\cite{MohazabAR08:bj}. 

MRSD has a simple intuitive
physical meaning-  the MRSD between two structures gives the average
distance each residue in one structure would have to travel on a straight line
to get to its counterpart in the other structure
(fig~\ref{fig:MRSDMeaning}).

\subsection*{Polymer non-crossing in protein folding}

\begin{figure}[hp]
  \centering
  \includegraphics[width=7cm]{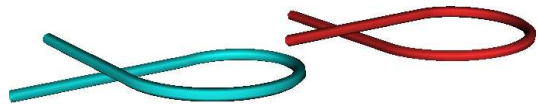}
  \caption{The MRSD and RMSD between the two curves are close to
    zero (the curves in this figure are displaced for better viewing but
    should be imagined to be superposed). But because the curve cannot
    pass through itself, in order to 
    undergo the transformation one leg must undergo relatively large
    amplitude motions to travel from one conformation to another. This
    results in a non-zero distance between the conformations by
    accurate metrics which can account for non-crossing. }
  \label{fig:slightOverUnder}
\end{figure}

The above interpretation of $MRSD$ points to a shortcoming of both $MRSD$
and $RMSD$, which is the importance of chain non-crossing
constraints. 
Consider the 
two curves depicted in fig~\ref{fig:slightOverUnder}, which
differ by having opposite sense of underpass/overpass. 
When both curves are aligned by minimizing $MRSD$ or $RMSD$, the
respective values are almost zero. However the physically relevant
distance for one conformation to transform to the other is much
larger, and must involves one arm of the backbone circumventing the
other as it moves between conformations. The transformation which
{\it minimizes} the distance  has been shown
previously to involve motions wherein one end of the polymer
doubles back upon itself until it reaches the underpass/overpass,
where it appropriately crosses under/over it, and
then proceeds snake-like to extend itself to the final
position~\cite{PlotkinSS07,MohazabAR08:bj}. 
We will not deal further with the aspects of non-crossing in this
paper.  

\section*{The generalized distance $\cal{D}$}

\begin{figure}[hp]
  \centering
  \includegraphics[width=5cm]{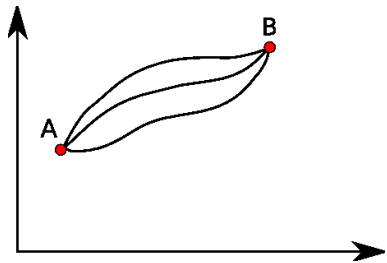}
  \caption{Distance between the two points $A$ and $B$ is the minimum
    length of the the curve connecting the two points.}
  \label{fig:distanceVariationalNotn}
\end{figure}

The distance between two points can be cast as a variational
problem, where the arclength of the curve between two
points is minimized (equation~\ref{eq:pointDistance}, see
fig~\ref{fig:distanceVariationalNotn}). 
The resultant Euler-Lagrange equations for the distance between two
points are:
\begin{equation}
  \label{eq:ELpoint}
  {d \over dt}({\partial {\cal L} \over {\partial \dot{\bf r}}})  = 0
\end{equation}
or
\[
\dot{\hat{\bf v}} = 0
\]
which means straight line motion, since this means that the direction of the velocity
does not change. 

As mentioned in the introduction, the notion of distance between two
points can be generalized to two curves or higher-dimensional objects in
general~\cite{PlotkinSS07}.
As in the case of points, the distance between two curves can be thought as a
variational problem, where one now minimizes the cumulative integrated
arclength between the two space curves:
\begin{equation}
  \label{eq:initD}
  {\cal D}[{\bf r}] = \int_0^L d s \int_0^T d t \sqrt{\dot{\bf r}^2}
  \: .
\end{equation}
Here  ${\bf r} \equiv {\bf r}(s,t) = (x(s,t), y(s,t), z(s,t))$ and $\dot{r}
\equiv \partial {\bf r}/\partial t$. 
The independent variables in this formulation are
position along the contour of the polymer $s$ and
elapsed ``time'' during the transformation $t$. 

\begin{figure}[hp]
  \centering
  \includegraphics[width=4cm]{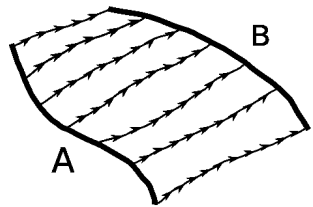}
  \caption{The distance ${\cal{D}}_{AB}$ is the accumulation of how
    much every part of the contour defining the space curve moves in
    the transformation between two conformations $A$ and $B$.}
  \label{fig:accumulatedDistance}
\end{figure}

Intuitively, the double integral in eq.~(\ref{eq:initD}) measures how
much every part of the polymer moves in going from one configuration
to another (see fig~\ref{fig:accumulatedDistance} for a schematic).

\begin{figure}[hp]
  \centering
  \includegraphics[width=5cm]{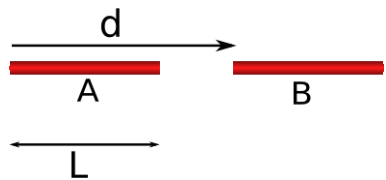}
  \caption{The line segment $A$ is displaced by $d$ along itself, to
    $B$. The soap film area $A_{soap}$ between the two segments is
    0. But the distance ${\cal D}_{AB} = L\,d$}
  \label{fig:noSoap}
\end{figure}

The minimal distance problem eq.~(\ref{eq:initD}) is not equivalent to
a simple soap-film problem (see fig~\ref{fig:noSoap}). It also has 
a lower symmetry than the relativistic   
world-sheet of a classical string~\cite{PlotkinSS07}, and so is
inequivalent to that problem as well. 

Minimizing equation~(\ref{eq:initD}) results in straight line motion
of all points along the curve. This is because
equation~(\ref{eq:initD}) models not an inextensible string but an
effective ``rubber band'' which can
expand and contract at no cost to facilitate the minimal-distance
transformation.  If the polymer cannot arbitrarily stretch and contract (a good
approximation for real polymers), the trajectories of the constituent
segments deviate from straight lines.

The polymer is made inextensible by introducing the constraint
\begin{equation}
  \label{eq:inextensible}
  \sqrt{\left({\partial {\bf r} \over \partial s}\right)^2} \equiv
  \sqrt{{\bf r}'^2} = 1  \: ,
\end{equation}
whereupon the function to be minimized becomes
\begin{equation}
  \label{eq:DLengthConstraint}
  {\cal D} = \int_0^L\int_0^T \!\!\! ds dt \: {\cal L}(\dot{\bf r}, {\bf r}')
\end{equation}
with effective Lagrangian:
\begin{equation}
  \label{eqL}
{\cal L}(s,t) = \sqrt{\dot{\bf r}} - \lambda(\sqrt{{\bf r}'^2} - 1) \: .
\end{equation}
and Lagrange multiplier $\lambda \equiv \lambda(s,t)$, a function of
both $s$ and $t$. 

The new equations of motion obtained by extremizing the functional become:
\begin{equation}
  \label{eq:newDEL}
  \dot{\hat{\bf v}} = \lambda {\bfk} + \lambda ' \hat {\bf t}
\end{equation}
where $\hat {\bf t}$ is the unit tangent vector and $\bfk$ is
the curvature vector~\cite{PlotkinSS07}.

\begin{figure}[hp]
  \centering
  \includegraphics[width=10cm]{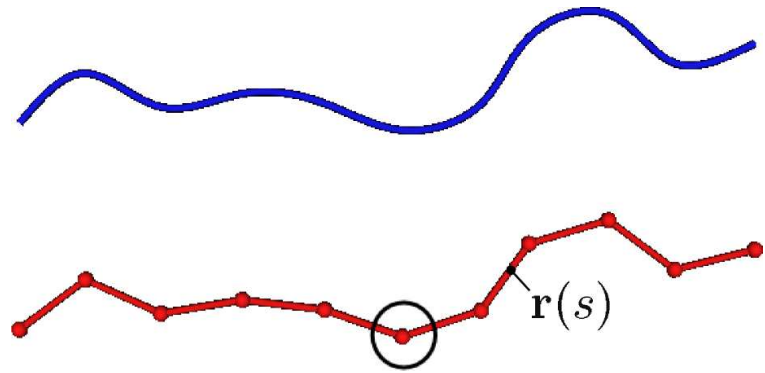}
  \caption{The lower curve is a discretized version of the upper
    one. After discretization the PDE for the upper curve becomes
    a set of $N$ coupled ODE's for the $N$ residues in the lower
    chain (A sample residue is marked with a circle).}
  \label{fig:discretization}
\end{figure}

Numerical solutions may be more readily obtained by discretizing the
string as shown in figure~\ref{fig:discretization}. 
This procedure is a particular example of the method of lines, used
to obtain solutions of partial differential equations. 
After discretization, the functional to be minimized becomes
\begin{equation}
  \label{eq:discretizedD}
  {\cal D}[{\bf r}_i, \dot{\bf r}_i] = \int_0^T dt {\cal L}({\bf r}_i,
\dot{\bf r}_i),
\end{equation}
where the effective Lagrangian $\cal L$ is:
\begin{equation}
  \label{eq:effL}
  {\cal L}({\bf r}_i,\dot{\bf r}_i) =\
  \sum_{i = 1}^N \left( \sqrt{\dot{\bf r}_i^2} - {\lambda_{i, i+1} \over 2}\
    \left(({\bf r}_{i + 1} - {\bf r}_i)^2 - b^2\right)\right).
\end{equation}
Here $b$ is the segment length which we set to unity. The distances we
obtain are thus in units of $b^2$. The distance between space
curves has the dimensions of area just as the distance between points has
dimensions of length. 
Upon discretization the PDE of the system becomes a set of $N$ coupled
ODE's, one for each residue:
\alpheqn
\begin{eqnarray}
  \label{eq:ELDiscr}
  \dot{\hat{\bf v}}_1 + \lambda_{12} {\bf r}_{2/1} & = & 0 \\
  \dot{\hat{\bf v}}_2 - \lambda_{12} {\bf r}_{2/1}  + \lambda_{23} \
  {\bf r}_{3/2} & = & 0\\
  \vdots &&  \nonumber \\
  \dot{\hat{\bf v}}_N + \lambda_{N-1,N} {\bf r}_{N/(N-1)} & = & 0  \:
  .
\label{eq:ELDiscrlast}
\end{eqnarray}
\reseteqn
The solutions of the first and last ($N$th) residues or beads consist of either
straight-line motion of the bead, pure rotation of the link
terminating on the bead, or a stationary solution where the residue remains at
rest. 
Moreover, {\it Weierstrass-Erdmann corner conditions} or 
{\it transversality conditions} demand smooth curves for solutions by
disallowing discontinuities or cusps in 
the trajectories~\cite{MohazabAR08}.

\begin{figure}[hp]
  \centering
  \subfigure[]{\includegraphics[width=5cm]{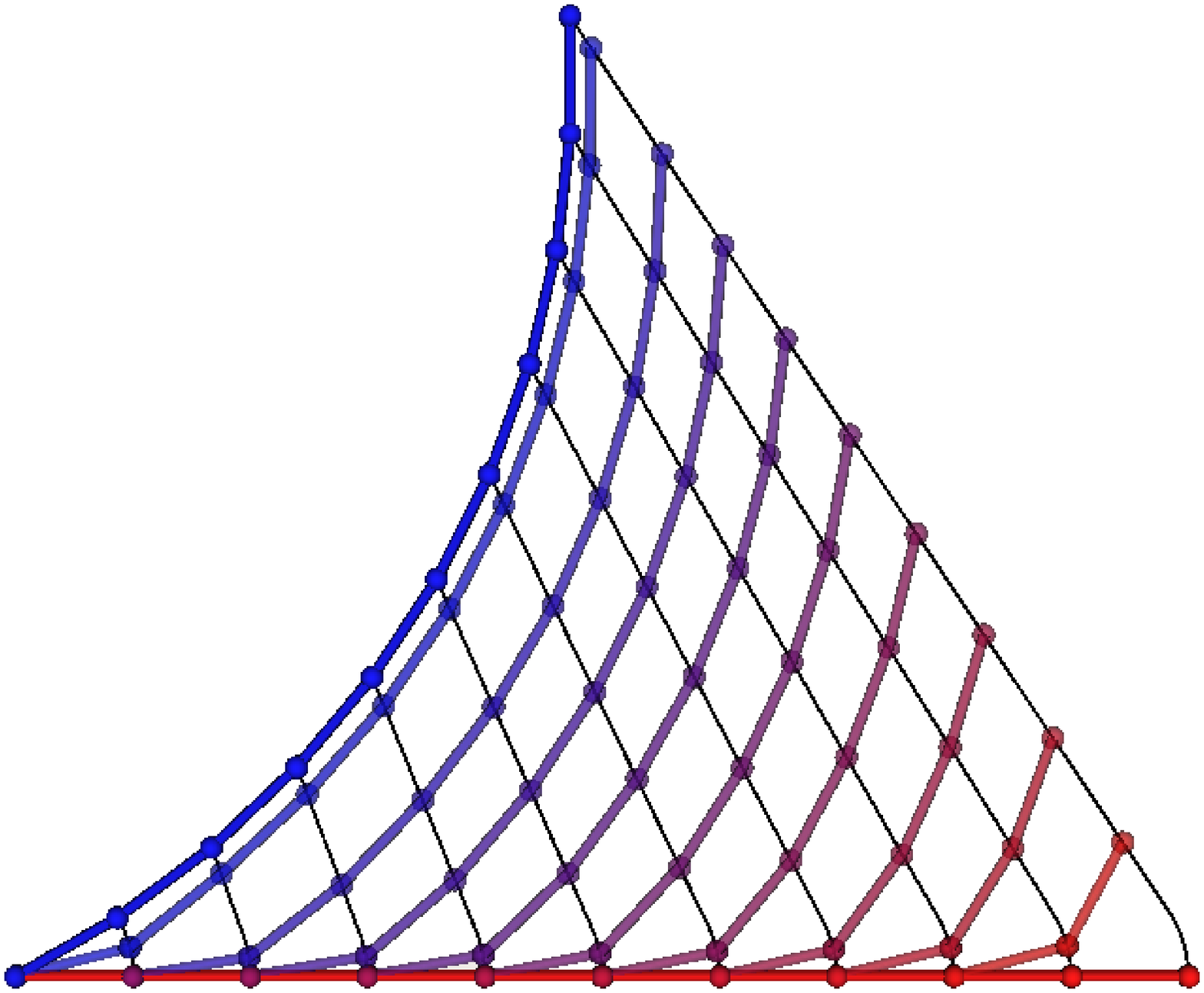}}
  \subfigure[]{\includegraphics[width=5cm]{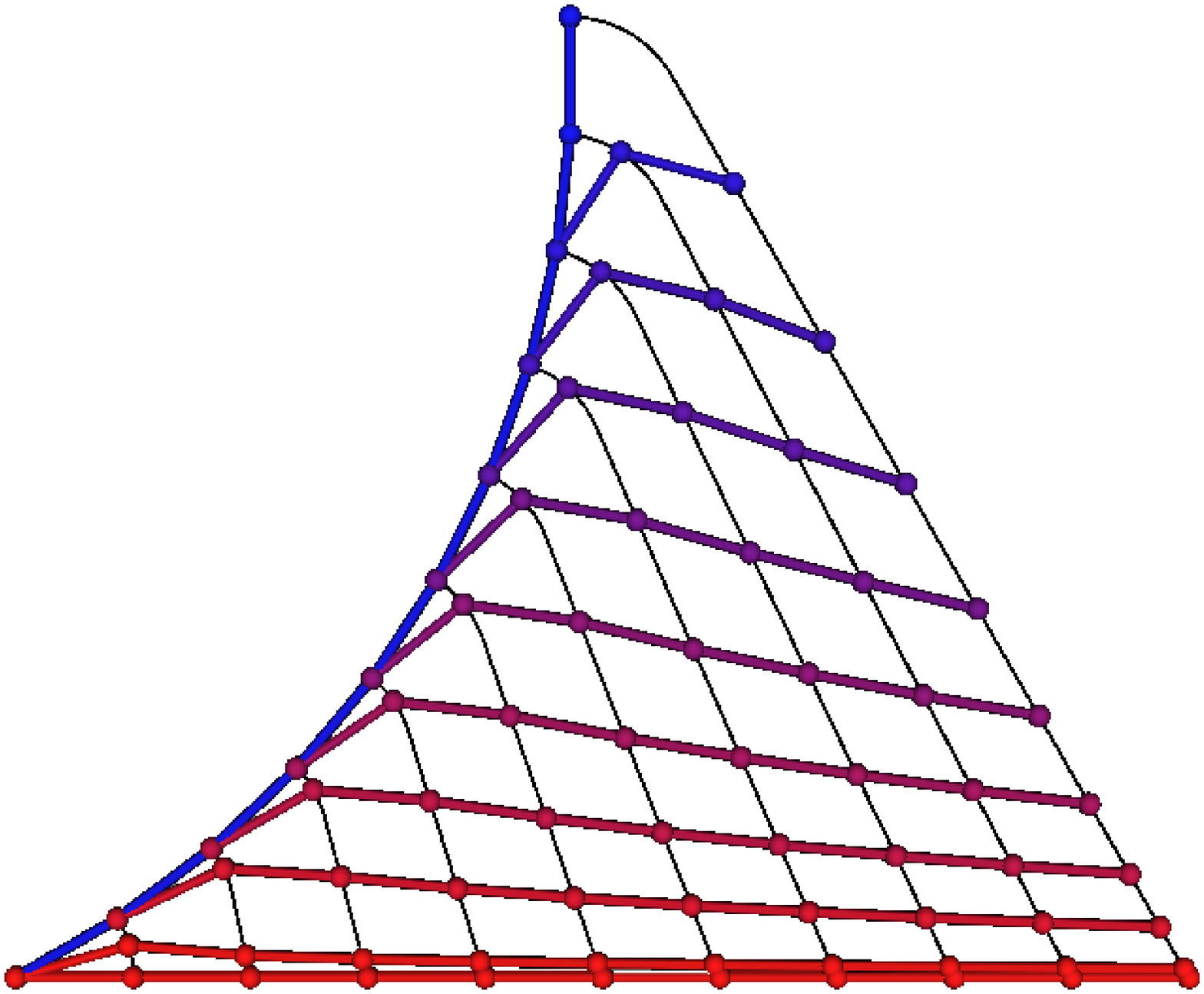}}
  \caption{Minimal and sub-minimal transformations between a straight
    line and a quarter circle (see text for description). For the left
    transformation ${\cal D} = 45.793$ and for the right one ${\cal D}
    = 46.278$}
  \label{fig:minimalSubminimal}
\end{figure}

Given two conformations which serve as boundary conditions on the equations of
motion~(\ref{eq:ELDiscr}-\ref{eq:ELDiscrlast}), several solutions
yielding slightly (non-extensively) different $\cal D$'s can be
constructed. It can be shown that they are all local minima~\cite{MohazabAR08}.
In figure~\ref{fig:minimalSubminimal} two solutions are shown. 
Figure~\ref{fig:minimalSubminimal}A depicts the global minimum
transformation, and figure~\ref{fig:minimalSubminimal}B a sub-minimal
``excited-state'' transformation.
The solutions both involve either rotations of
the constituent links or 
straight line motion of the constituent beads. 
In figure~\ref{fig:minimalSubminimal}A, rotation occurs away from the
straight-line conformation and results in a distance ${\cal D} =
45.793$, while in~\ref{fig:minimalSubminimal}B 
rotation occurs from the curved conformation and results in ${\cal D}
= 46.278$.

The fact that a real polymer cannot cross itself can be incorporated
into the problem of 
finding the minimal distance~\cite{MohazabAR08:bj}. 
Non-crossing is manifested as an inequality
constraint~\cite{PontryaginLS62,CassD65,GregoryJ94,GregoryJ07},
which appears 
in equation~(\ref{eq:effL}) as a Lagrange parameter for each residue
$i$, multiplying the excluded volume constraint. To describe this, let
the unit vector from the $k$th to the $(k+1)$th bead be $\hat{e}_k \equiv
(\bfr_{k+1} - \bfr_{k})/b$, then the vector to position $\bfr(s)$ at
contour length $s$ on the chain (see
e.g. fig~\ref{fig:discretization}) is 
\bea
\bfr(s) &=& b \sum_{i=0}^{k-1} \hat{e}_i + (s-k b) \hat{e}_k \nonumber \\
&=& \bfr_k + (s - k b) \hat{e}_k \nonumber \: . 
\eea
To constrain the motion of the beads so that the chain cannot cross
itself, we add the term
\be
\l_i \left( \int_0^s \!\! ds \: \left| \bfr(s) - \bfr_i \right| +
  \eps_i^2 \right)  
\label{eqnc}
\ee
to the summand of equation~(\ref{eq:effL}). Note that by discretizing
the problem to find the motion of residues, there must be an asymmetry
in the way that the chain is treated-- in a continuum treatment the
term in the integrand of~(\ref{eqnc}) would be $| \bfr(s) -
\bfr(s')|$. The quantity $\eps^2$ in~(\ref{eqnc}) is an ``excess
parameter'' which is zero unless a residue is directly constrained
(touching some part on the rest of the chain). If $\eps_i^2=0$ the problem of
finding minimal distance is a ``free'' problem for residue $i$, and the equations of
motion~(\ref{eq:ELDiscr}-\ref{eq:ELDiscrlast}) are unchanged. However
the corner conditions mentioned above induce an implicit ``knowledge'' of
the sterically avoided boundary, so that the motion of the residues
are altered to travel most directly to the steric surface constituting the
constraint or obstacle.  At this point the
residue is constrained to be on the surface of the obstacle and the
trajectory is defined accordingly. 
Subsequently the residue leaves the constraining 
surface and the problem becomes a free problem once
again, travelling most directly to the final
conformation~\cite{MohazabAR08:bj}. 

In the above treatment the chain has zero thickness. A tube thickness
$\rho$ can
be straightforwardly incorporated into the treatment by letting $\bfr(s) \rightarrow
\bfr(s) + \rho \hat{e}_{\rho}$ in equation~(\ref{eqnc}), and then
integrating over the surface of the cylinders which compose the
resulting piece-wise tube. 

Another modification that can be made to the Lagrangian is one
involving the curvature constraints. In the current treatment the
angle between to consecutive links of the chain can have any value,
whereas in real protein chains angles defined by bonds between atoms
or residues are restricted. We will not discuss these aspects in this manuscript.

\subsection*{The minimal distance between protein fragments}

In ref.~\cite{MohazabAR08:bj} protein fragments such as an alpha helix and
beta hairpin were considered for purposes of calculating the minimal
distance. An extended strand was aligned to the respective structures
by minimizing either RMSD or MRSD, and the distance $\cal D$ was
subsequently calculated for the aligned structural pairs. Both real
and idealized protein fragments were considered. 
Most pairs of structures had smaller distance minimal pathways when
aligned using MRSD as the cost function. In some 
cases however the smaller distance minimal pathway was obtained when the boundary
conformations were aligned using RMSD as the cost function.

\begin{figure}[hp]
  \centering
  \subfigure[]{\includegraphics[width = 2.5cm]{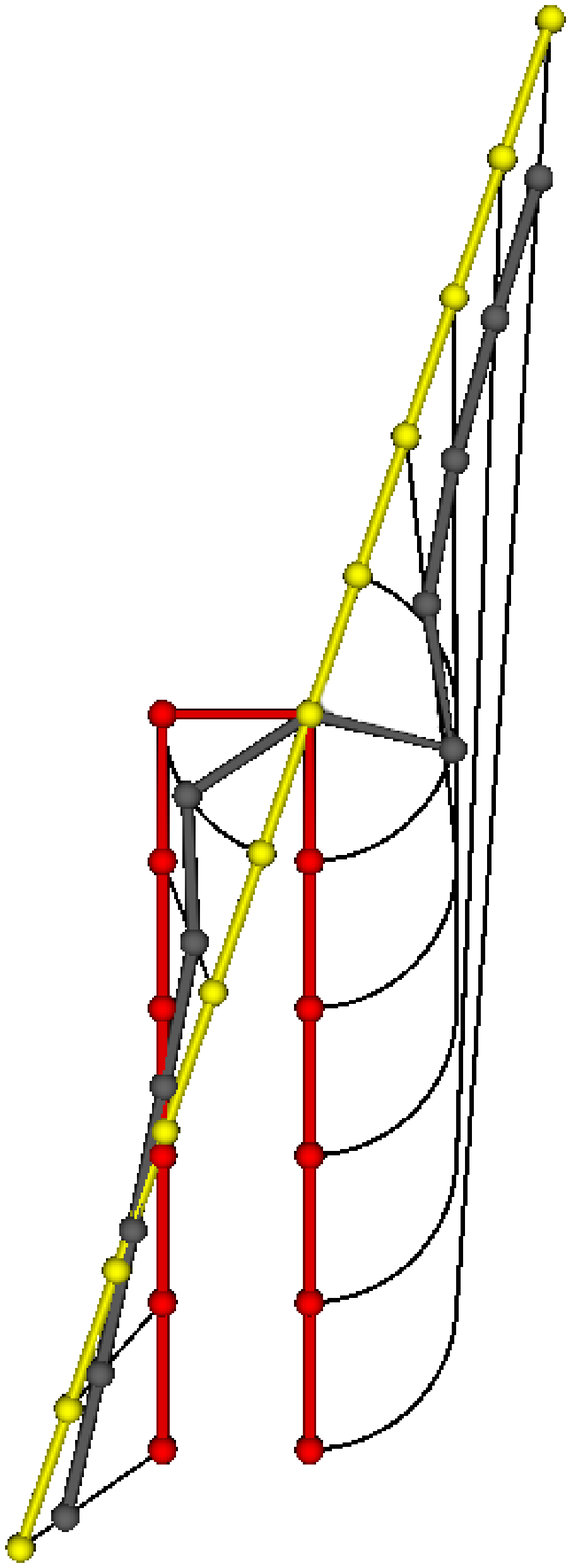}}
  \subfigure[]{\includegraphics[width = 7cm]{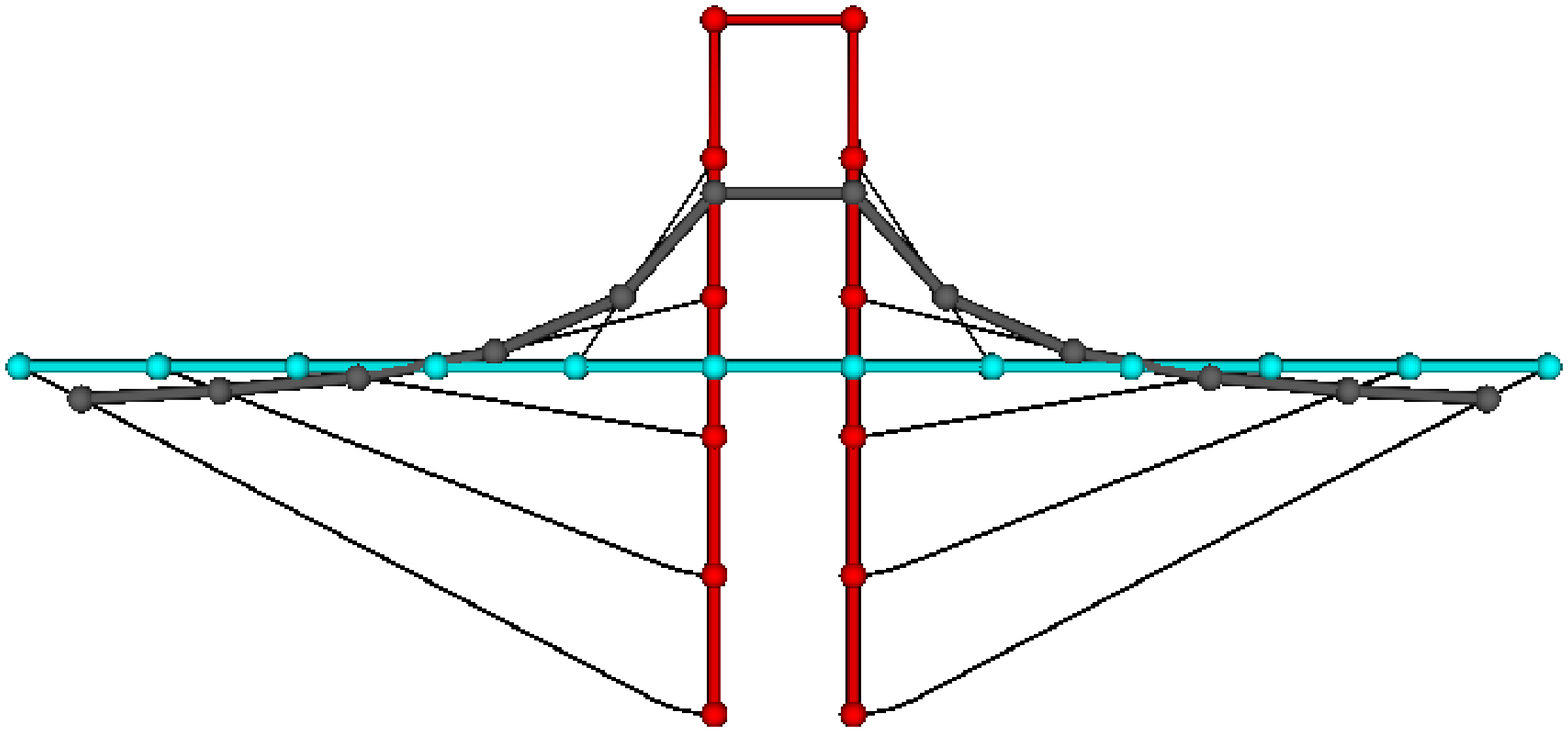}}
  \caption{(color) $\cal D$ minimizing transformations for MRSD aligned
    (yellow) and RMSD (cyan) aligned hairpins. Intermediate state is
    shown in grey. The distances for each transformation, in units of
    link length squared, are $3.20$ for MRSD-aligned and $3.22$ for
    RMSD-aligned structures.}
  \label{fig:SampleIBTransformation}
\end{figure}

For example, the straight line conformation in
figure~\ref{fig:SampleIBTransformation} was aligned to an idealized
$\beta$-hairpin structure also shown in that figure. 
The alignment was performed by both
minimizing the MRSD between the structures
(figure~\ref{fig:SampleIBTransformation}A), and by minimizing
RMSD between the structures
(figure~\ref{fig:SampleIBTransformation}B). 
In each instance, the minimal 
distance $\cal D$ between the structural pairs was calculated after
alignment.  The resulting aligned straight-line structures have significantly
different position/orientation depending which cost function was used, MRSD or
RMSD: the MRSD between the two staight-line structures is in fact
larger than the MRSD between 
each and the hairpin structure~\cite{MohazabAR08:bj}.

Both transformations are minimal transformations but are subject to
different boundary conditions and thus yield different pathways and
${\cal D}$'s. The question remains as to how to align the structures to
obtain the minimum of all minimal transformations, i.e. the minimum 
minimal distance ${\cal D}$.  
To calculate this quantity, ${\cal D}$ itself must be used as the cost
function for alignment.\footnote{In the limit of a large number of
  residues ($N$), the distance 
  converges to the $N$ times the $MRSD$: ${\cal D} \rightarrow N\times
  MRSD$, so for long chains MRSD can be 
  considered a first step towards optimal alignment. But ideally one
  wants to align the two structures using $\cal D$ itself as a cost
  function.} 

In this paper, we align structures using ${\cal D}$
as a cost function to obtain for the first time the minimum of all
minimal transformations. 
The structures that we consider are idealized straight-line segments with
varying number of links,  which are then aligned to idealized beta hairpins
using ${\cal D}$ as a cost function. The alignment and resulting
distance ${\cal D}$ are compared with the alignments and distances of
RMSD and MRSD. This is a first step toward aligning more complex
structures using $\cal D$ as a cost function. We will also see that
there exist high order approximations which capture much of the
properties of a true $\cal D$ alignment. Applying these approximate metrics 
to align structures such as a full protein is a topic for future research.

\section*{Structural alignment of protein fragments using the distance
$\Dist$}

In principle, minimal
pathways can be computed for any initial and final configurations,
just as RMSD can be computed between any two configurations. 
However it of special significance to anneal the configurations
allowing translations and rotations, until the minimal distance
transformation is achieved (i.e. the minimum of minimal distance
transformations). This is analogous to the usual procedure of using 
RMSD or MRSD as a cost
function between two structures and minimizing with respect to
translations and rotations. While the minimization procedure is particularly
straightforward for RMSD and involves the inversion of a matrix, the
minimization using the distance $\Dist$ as a cost function involves 
a simplex or conjugate gradient minimization and so is more 
computationally intensive. 

In short the boundary conformations are allowed to translate and
rotate in 3D space. Their position and
orientation is modified to produce a pathway with minimal length, as compared
to all other minimal pathways that can be obtained by positioning and orienting
the same two structures in 3D space.

\subsection*{Method and Results}

For the purpose of generating accurate initial guesses for the minimal distance
aligned structure, we introduce the following hierarchy:
\alpheqn
\begin{align}
  \label{eqhier}
  {\cal D}_0 &= N \times MRSD &\includegraphics[width=3.5cm]{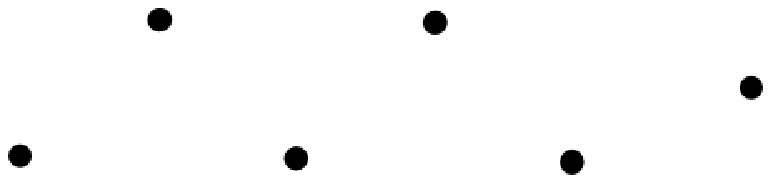}
\\
\label{eqd1}
{\cal D}_1 &= \sum_{i=1}^{N-1} {\cal D}\left(
\ell_i^{\left(\mbox{\tiny A}\right)}, \ell_i^{\left(\mbox{\tiny
      B}\right)} \right) &\includegraphics[width=3.5cm]{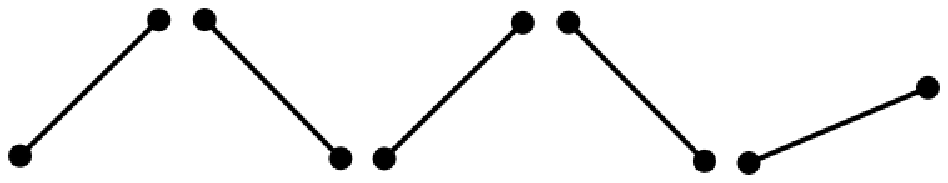}
\\
{\cal D}_2 &= \sum_{i=1}^{\mbox{\scriptsize{int}}\left((N-1)/2\right)} {\cal D}\left(
\left\{ \ell_i^{\left(\mbox{\tiny A}\right)} \right\} , 
\left\{ \ell_i^{\left(\mbox{\tiny B}\right)} \right\}
\right)  + {\cal D}_1^{\left(\mbox{\tiny{end link}}\right)} &\includegraphics[width=3.5cm]{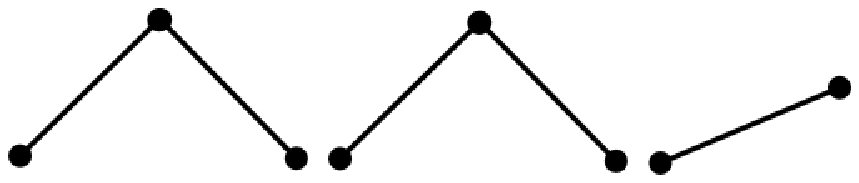} \\
\vdots& & \nonumber   \\
{\cal D}_N &= {\cal D}\: . &\includegraphics[width=3.5cm]{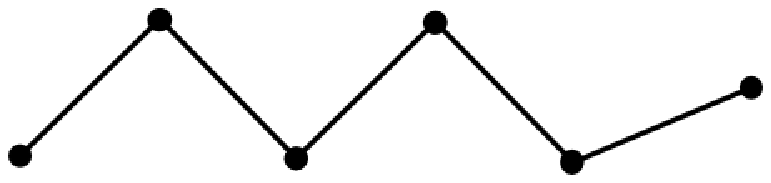} 
\end{align}
\reseteqn

In this hierarchy, the ${\cal D}_{\alpha}$ have the following
interpretation: ${\cal D}_0$ is the cumulative distance between the
sets of points comprising the residue locations of conformations $A$
and $B$, ${\cal D}_1$ is the cumulative distance between the sets of
single links, $\ell_i$, comprising configurations $A$ and $B$, ${\cal
  D}_2$ is the cumulative distance between the sets of double links,
$\left\{\ell_i\right\}$, comprising configurations $A$ and $B$ plus any
single-link remainder if one exists, and so on. That is, at level $\alpha$ the polymer
chain is divided up into sub-segments each of link-length $\alpha$,
plus one segment constituting the remainder. When $\alpha=N$, the
chain as a whole is considered, which is the true distance ${\cal D}$.
This procedure is also illustrated schematically adjacent to each
equation above. 

We observed that ${\cal D}_1$ was a good approximation to the total
$\cal D$ between two chains, was much easier in practice to
calculate, and could be automated in a robust way. For these reasons we
used it to generate initial guesses for minimal distance aligned
structures.  After the initial alignment using ${\cal D}_1$ the chains were further
aligned using the full distance $\cal D$. At this stage the general form
of the transformation is established and the computation can be
automated.  We used a Nelder-Mead simplex method in our algorithm to find the
minimal distance alignment.

\begin{figure}[hp]
    \subfigure[]{\includegraphics[height =1.125cm, angle=270]{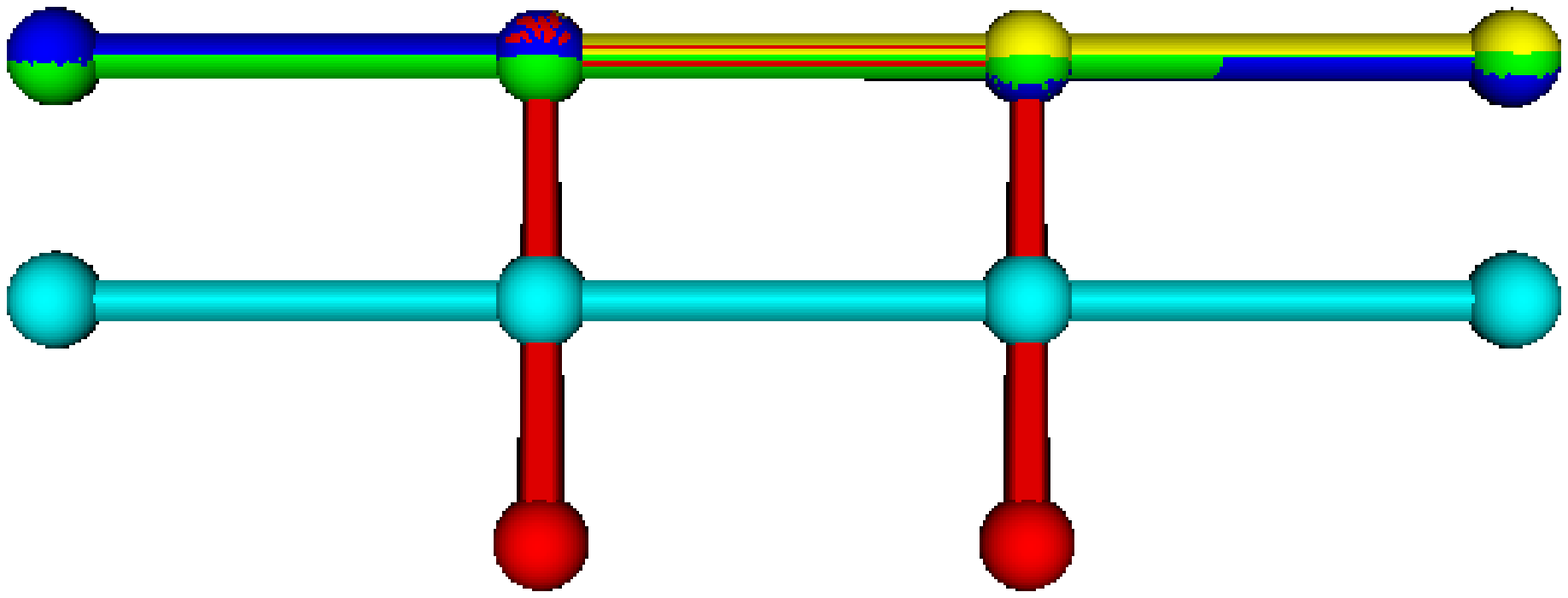}}
    \subfigure[]{\includegraphics[height =1.875cm, angle=270]{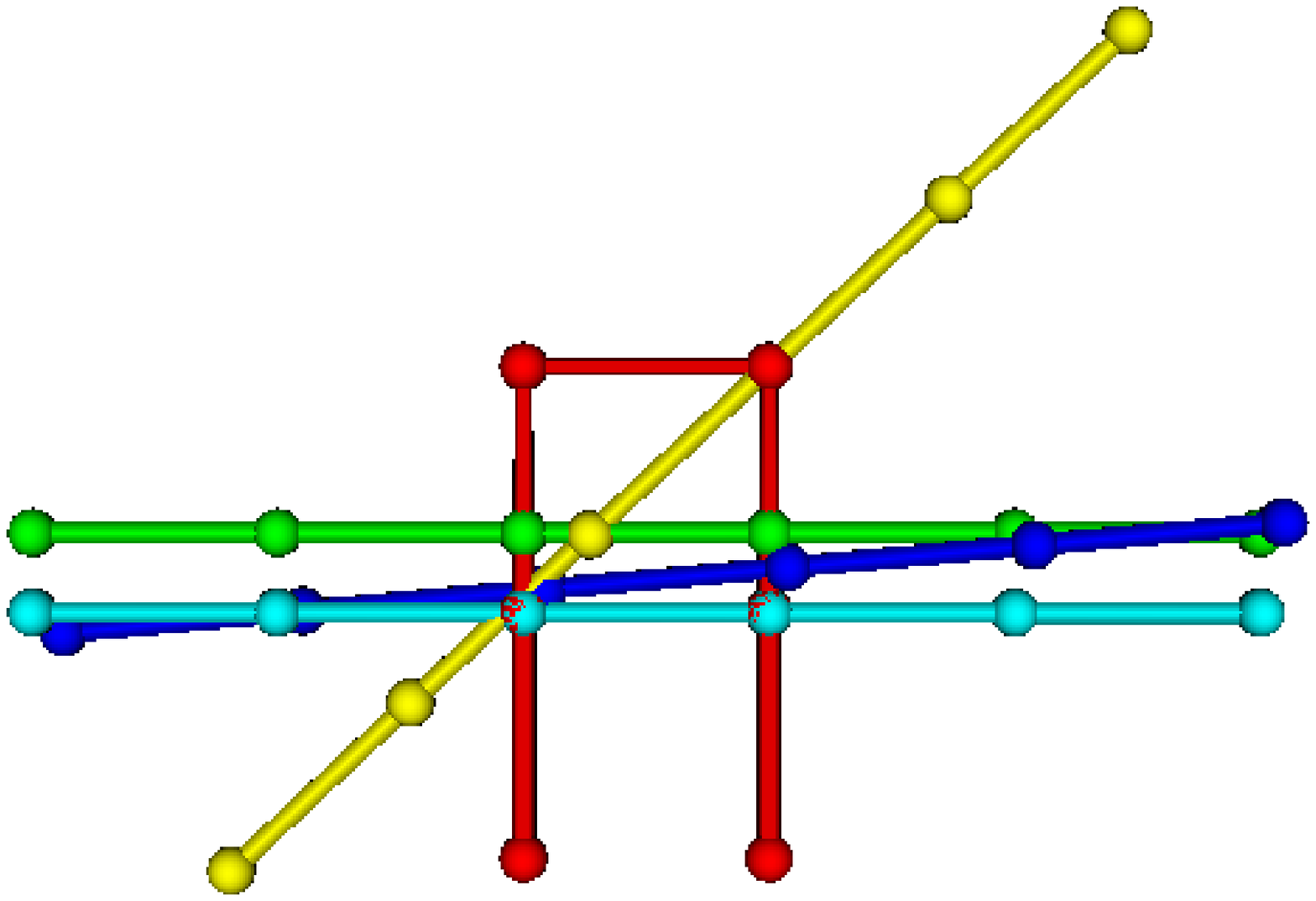}}
    \subfigure[]{\includegraphics[height =2.625cm, angle=270]{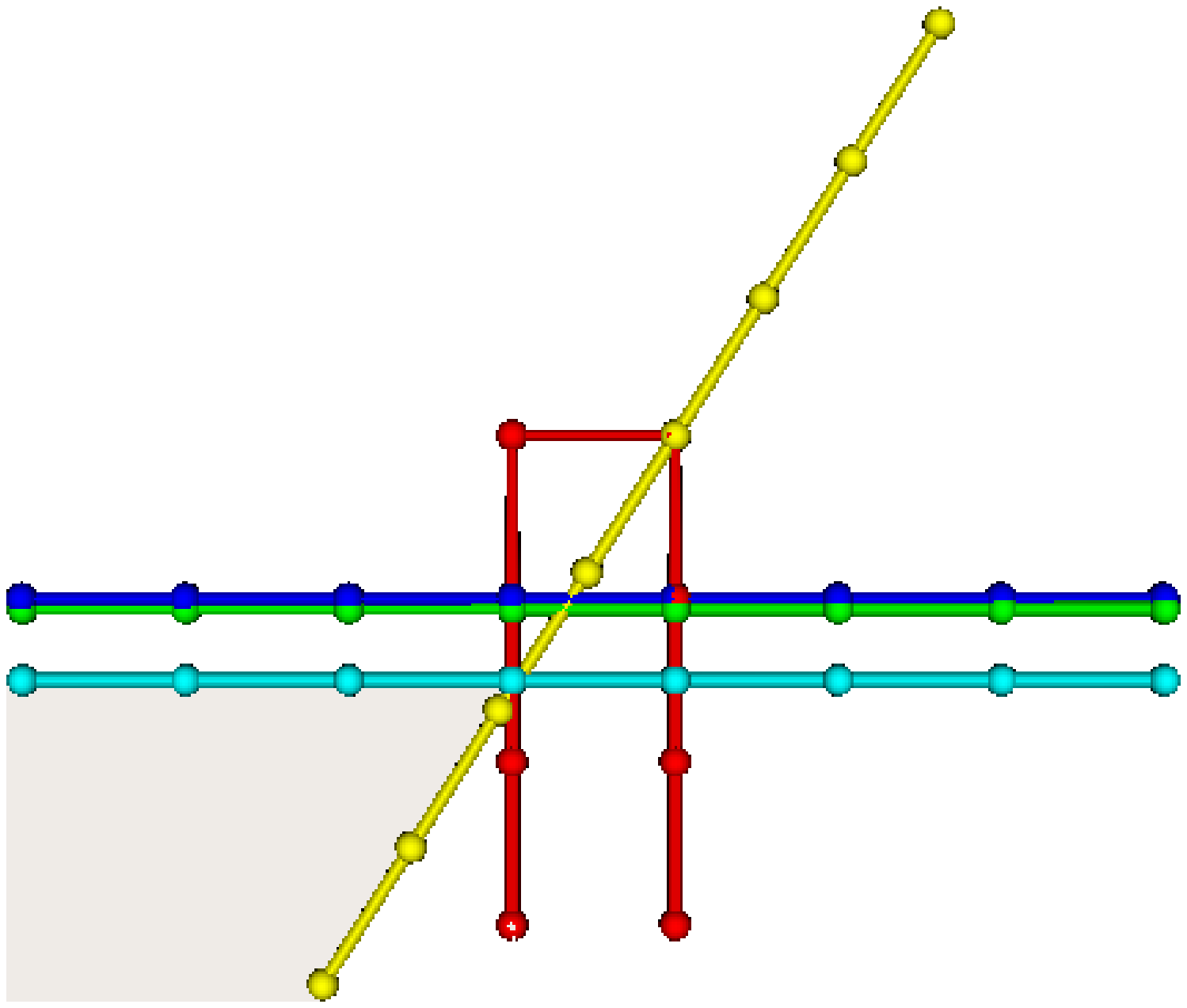}}
    \subfigure[]{\includegraphics[height =3.375cm, angle=270]{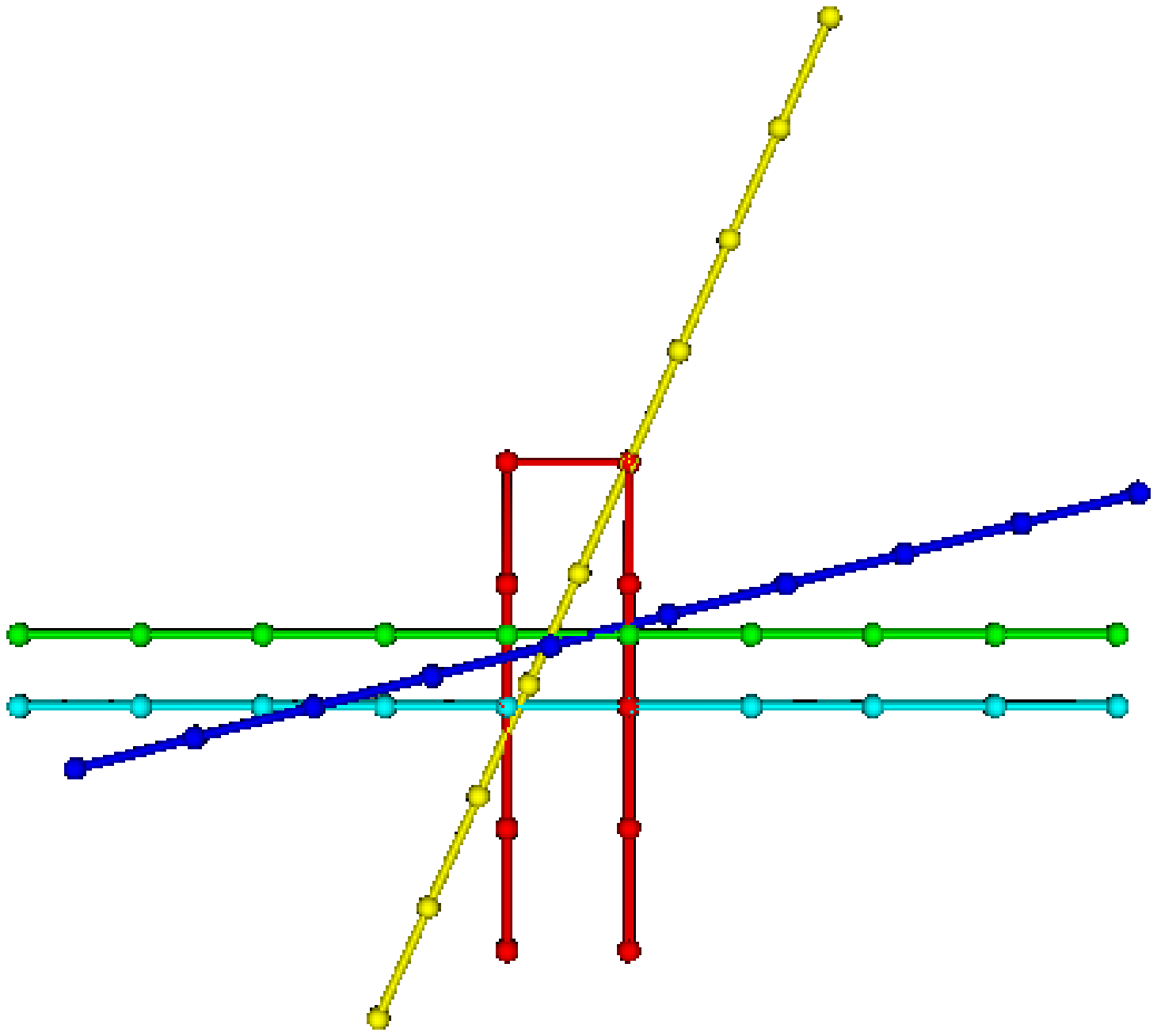}}
    \subfigure[]{\includegraphics[height =4.125cm, angle=270]{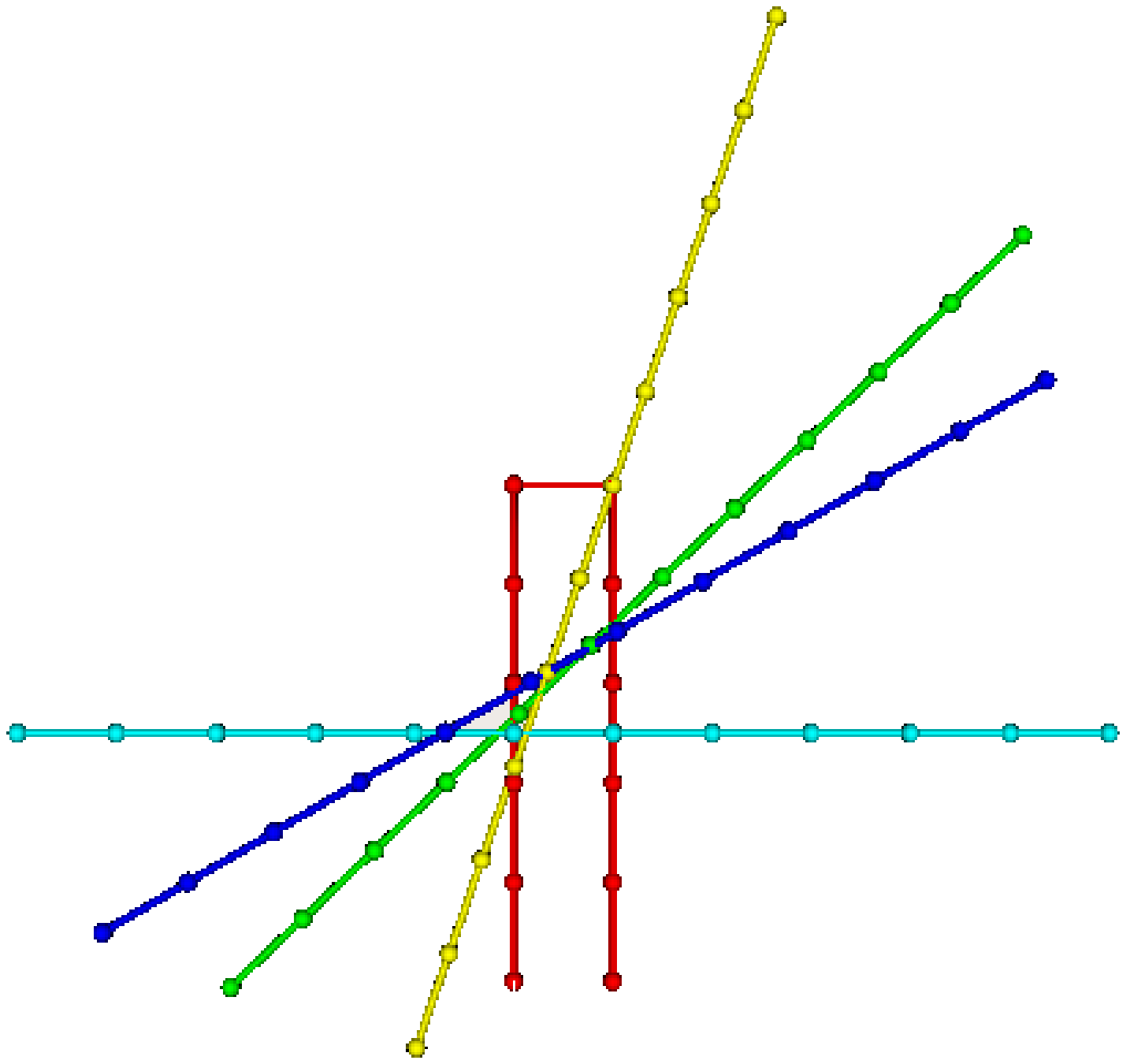}}
    \subfigure[]{\includegraphics[height =4.875cm, angle=270]{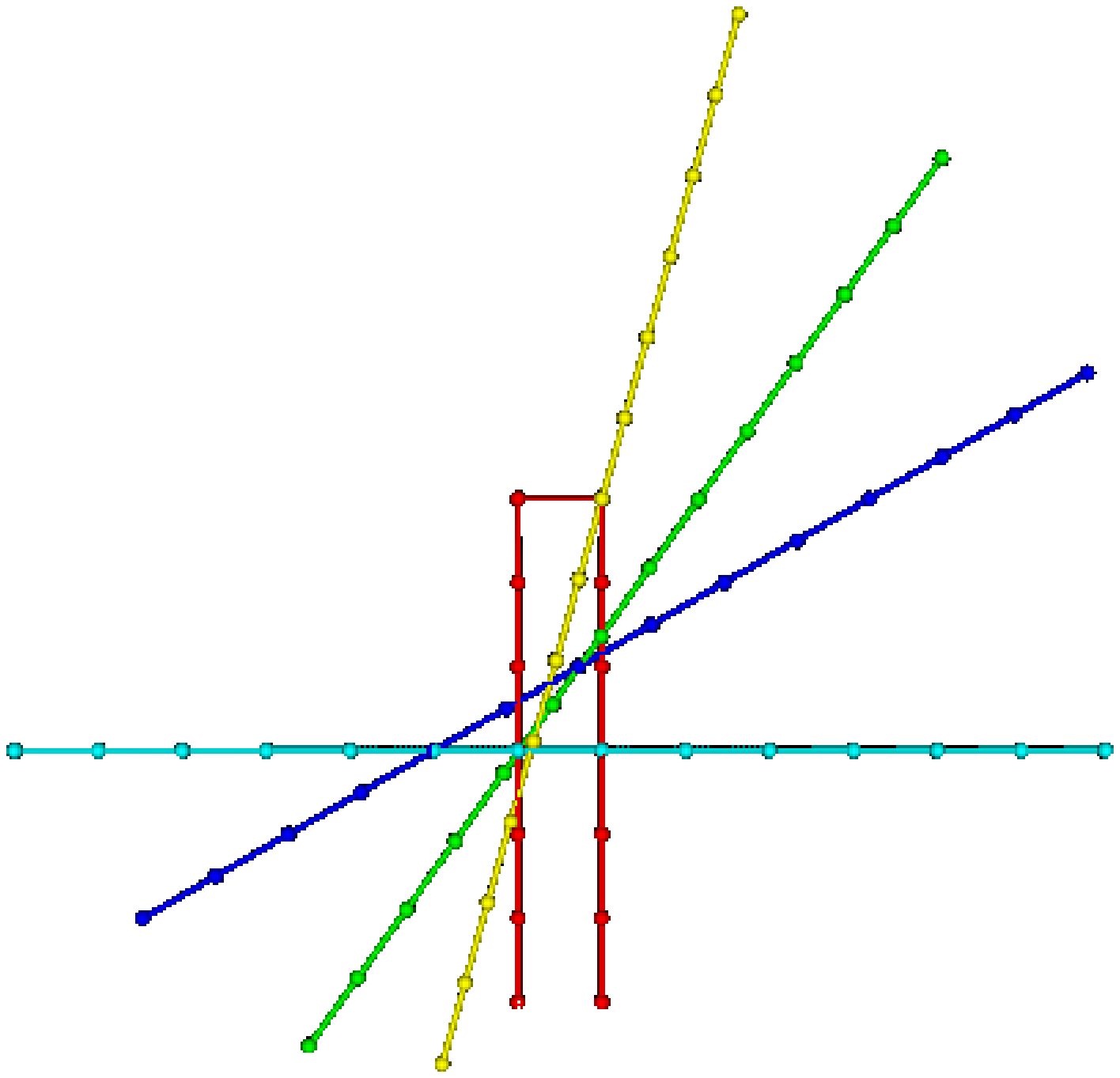}}
    \subfigure[]{\includegraphics[height =5.625cm, angle=270]{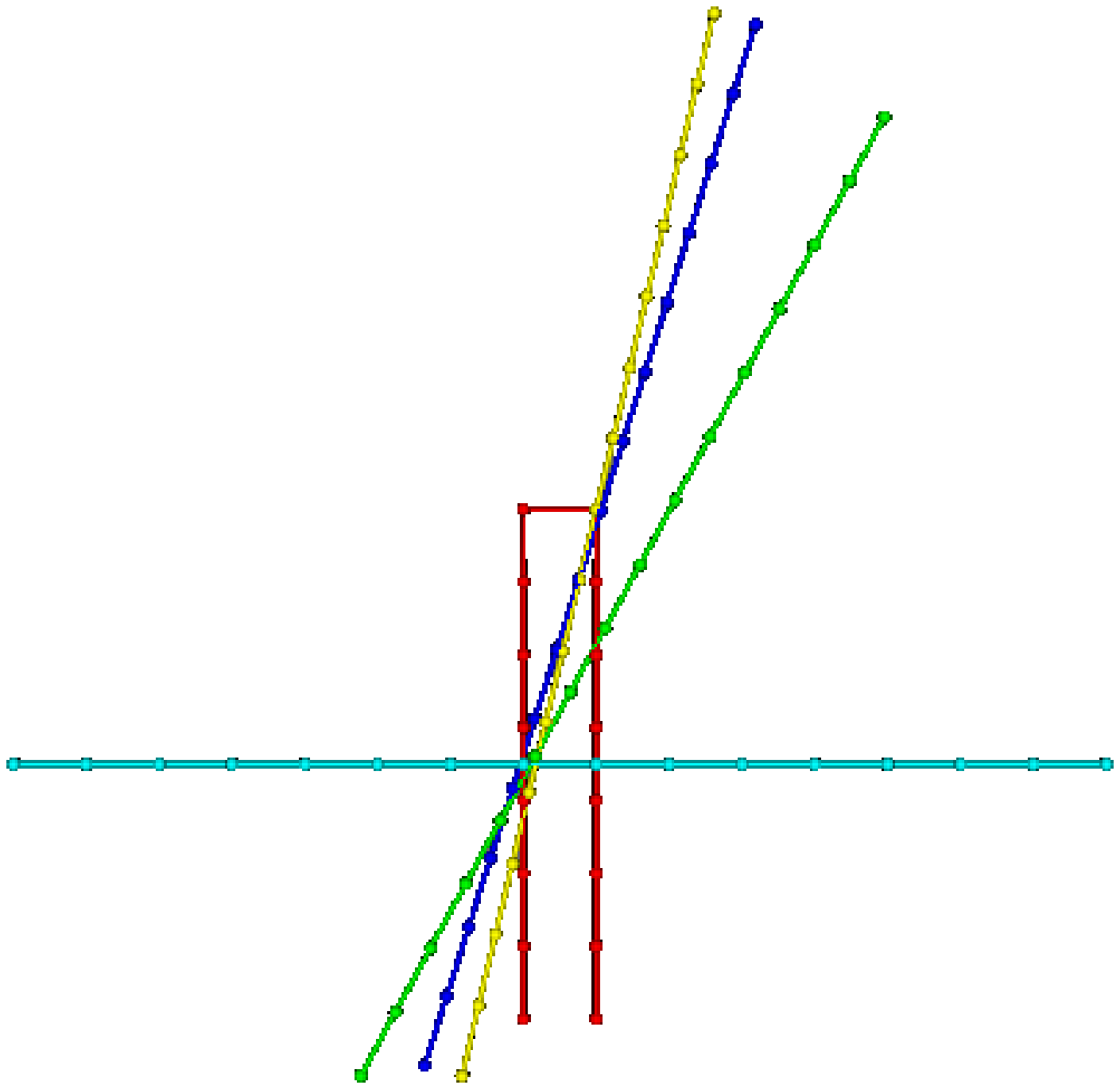}}
    \subfigure[]{\includegraphics[height =6.375cm, angle=270]{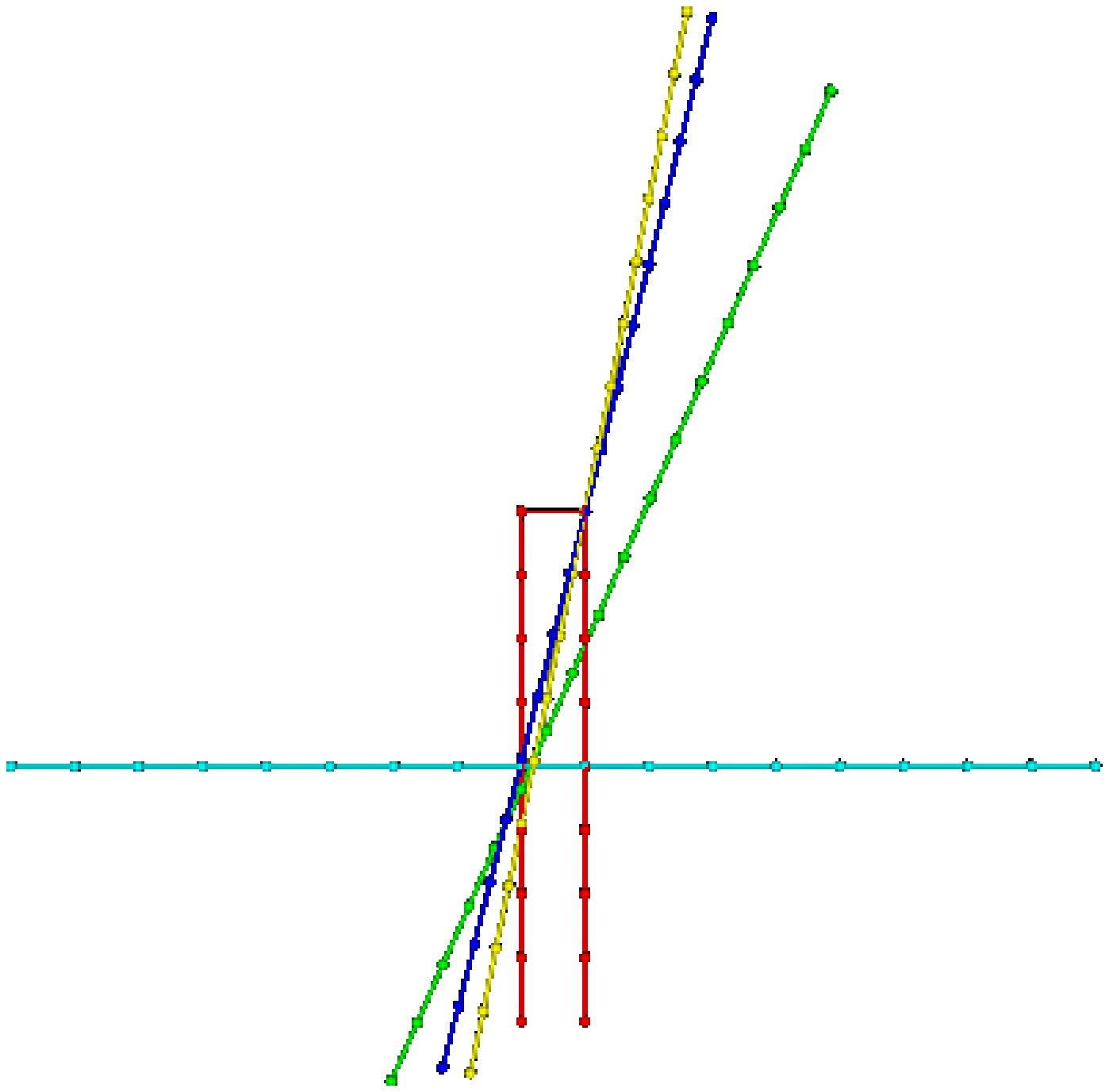}}
    \subfigure[]{\includegraphics[height =7.125cm, angle=270]{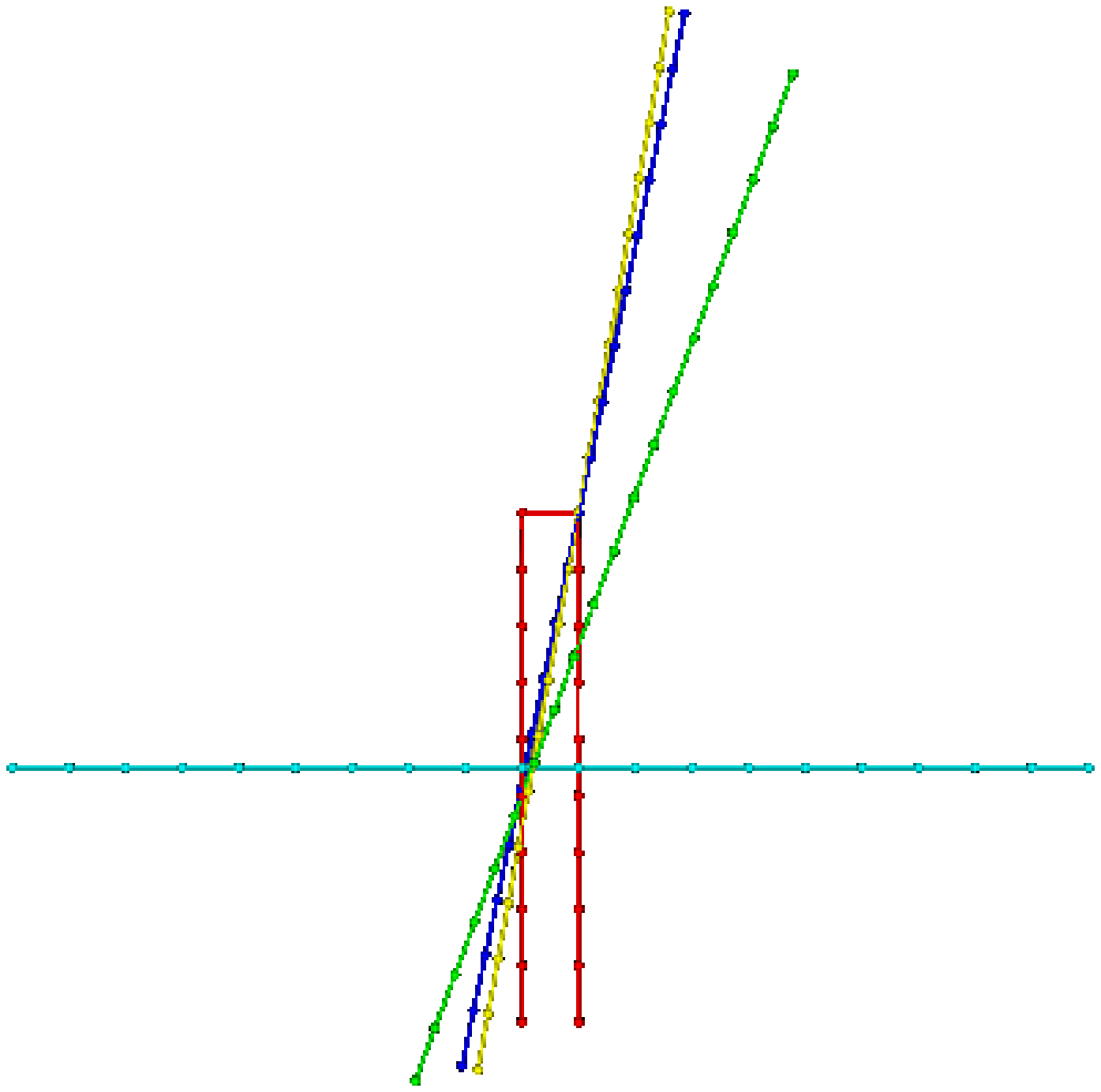}}
    \subfigure[]{\includegraphics[height =7.875cm, angle=270]{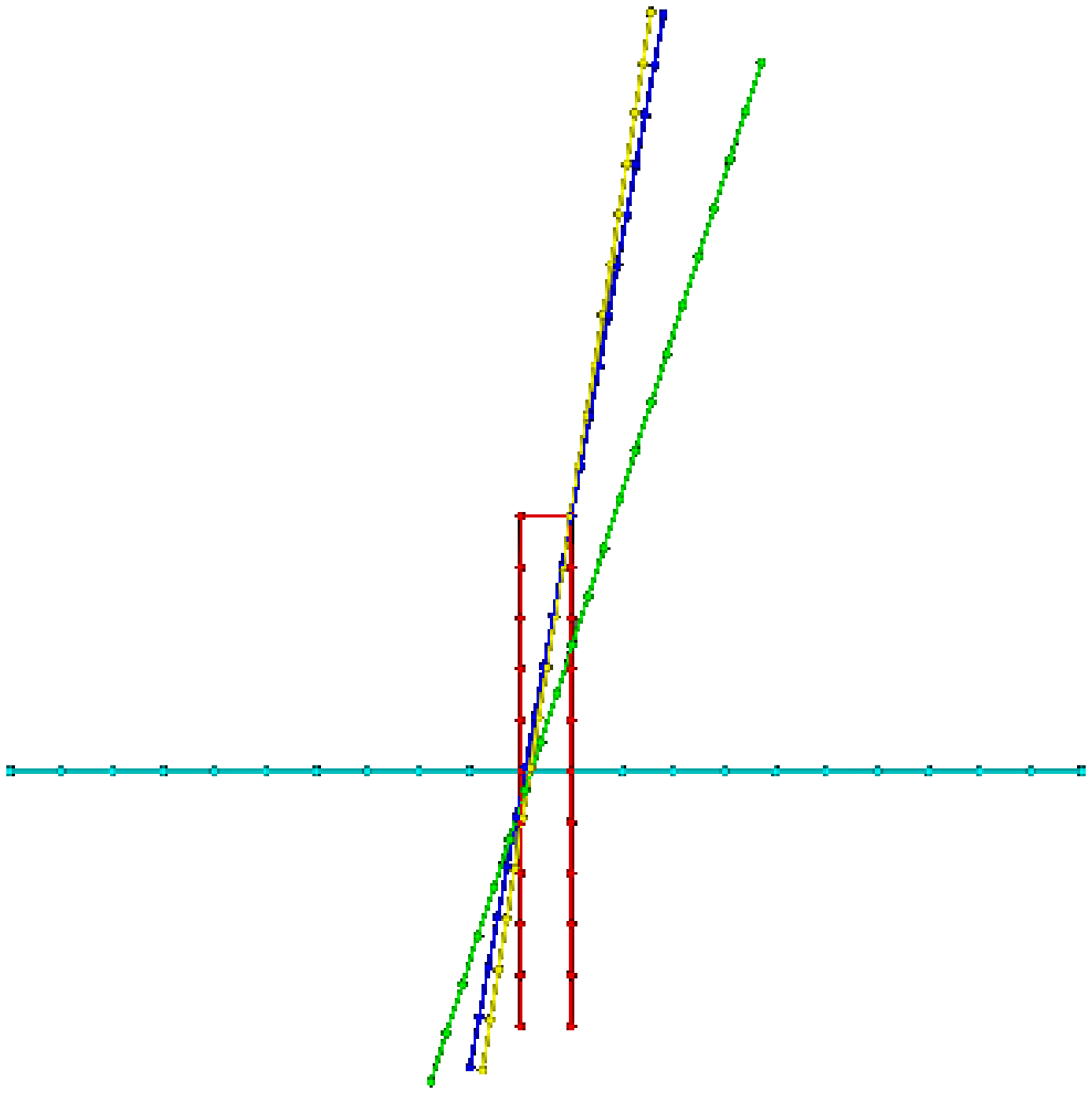}}
  \caption{(color) Alignments with different cost functions. The Hairpin is
    shown in red. $\cal D$ alignment in green, ${\cal D}_1$ in blue,
    MRSD in yellow, and RMSD in cyan}
  \label{figalignments}
\end{figure}

Figure~\ref{figalignments} shows the aligned structures using $RMSD$,
$MRSD$, $\Dist_1$, and $\Dist$, for increasing numbers of
links. Several points can be observed. For the smallest number of
links ($3$), $MRSD$, $\Dist_1$, and $\Dist$ all give the same
alignment (fig~\ref{figalignments}a). For $5$ or more links, the
$MRSD$-aligned structure breaks 
symmetry by choosing particular diagonal direction, while $\Dist_1$
and $\Dist$ retain this symmetry but begin to differ (fig~\ref{figalignments}b). 
The deviation from $MRSD$ and $\Dist$ is a finite-size
effect~\cite{PlotkinSS07}, so we 
know that the two alignments must eventually converge as $N$ is
increased. At $9$ links (fig~\ref{figalignments}d), the $\Dist_1$-alignment breaks
symmetry in the same fashion as MRSD, yet the $\Dist$-alignment
remains similar to RMSD. By $11$ links (fig~\ref{figalignments}e),
the $\Dist$-aligned structure has broken symmetry as well, however
with a smaller angle to the horizontal than either $MRSD$ or
$\Dist_1$. As $N$ is increased, $\Dist_1$ and $MRSD$ aligned
structures quickly converge,
while the angle with respect to the horizontal of 
the $\Dist$-aligned structure continues to lag behind that of either
$MRSD$ and $\Dist_1$ structures,  converging slowly as $N$ continues to
increase (figures~\ref{figalignments}f-j). The $RMSD$-aligned
structure remains horizontal throughout. 

Average lengths of $\beta$-hairpins in databases constructed from the
PDB are about $17$ residues~\cite{CruzX02}, most consistent with
fig~\ref{figalignments}h. 
From this figure we see that hairpins of this length have
a globally different structural alignment with extended structures
depending on whether $\Dist$ or $RMSD$ is used.

\begin{figure}[hp]
  \centering
  \includegraphics[width = 10cm]{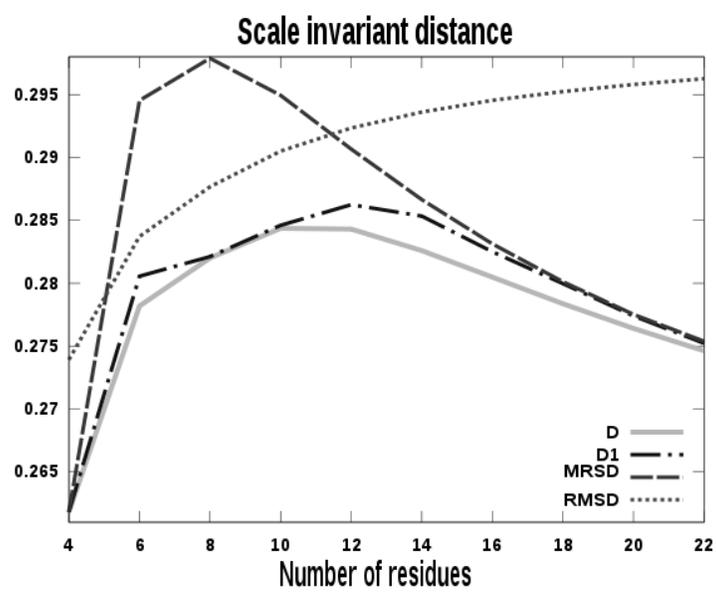}
  \caption{Scale invariant distance resulting from different alignments with different cost functions}
  \label{figscaleinvard}
\end{figure}

Table \ref{tab:results} and figure~\ref{figscaleinvard} summarize the
results for the minimal 
distance transformations from the aligned structures. 
Table~\ref{tab:results} gives the numerical value of the 
distance $\Dist$ for each aligned structure, aligned using the various
cost functions listed: $\Dist$, $\Dist_1$, $MRSD$, and $RMSD$.  
Note that the distance $\Dist$ is always
minimized for the distance-aligned structure, and tends to increase as
one considers the $\Dist_1$, $MRSD$ and then $RMSD$-aligned
structures for a given number of links. 

For comparison, in table~\ref{tab:resultsmrsd} the corresponding
values of MRSD are given for the aligned structures using each cost
function.  Note in each table that as $N \to \infty$,  $\cal D$
tends to converge to MRSD.

The distance travelled per residue, in units of link length is
$\Dist/N b$. Dividing this measure by the chain length $(N-1) b$ gives
a scale-invariant measure of the distance: $\tilde{\Dist} = \Dist/(N
(N-1) b^2)$. This quantity is plotted in
figure~\ref{figscaleinvard}. We can see from the plot that the
$D_1$-aligned structure generally gives a good approximation to the
true $\Dist$-aligned structure. Moreover, $MRSD$, $\Dist_1$ and
$\Dist$ all converge to the same while RMSD converges to a dissimilar
value.

\section*{Conclusion and Discussion}

In this paper, we reviewed the concept of the generalized distance
$\cal D$, and then used it as a cost function to align unfolded idealized
strands of various sizes to their corresponding idealized
$\beta$-hairpin structures. This is the first time
that the true Euclidean distance has been used as a cost function for
structural alignment. 
The distance $\cal D$ for the minimal
transformation between aligned structural pairs was compared for
various alignment cost functions: $RMSD$, $MRSD$, ${\cal D}_1$, and
$\Dist$ itself. $\Dist_1$ is the distance between conformational pairs
if the chain were decimated to single links and distance of all
single-link transformations was summed.

We found that $\Dist_1$-aligned structures generally gave a distance
that was close to the true $\Dist$-aligned structure, and in this
sense was a good approximation. However the aligned structures were
noticeably different depending on the cost function, for the finite
values of $N$ that we studied. Our largest value of $N$ was $22$
residues, while the average length of $\beta$-hairpins is about $17$
residues. 
For these average hairpin lengths, the minimal $\Dist$ aligned structure is
globally different from the $RMSD$ structure. 
Whether this discrepancy is generally
true for larger structures or whole proteins remains to be determined,
but we feel it is likely. It is not yet clear at this point whether
alignment using distance will yield more accurate predictions for such
problems as 
protein structure prediction or ab-initio drug design. What is clear
is that the best-aligned structures using a reasonable alignment
metric such as the true distance give very different results than
RMSD, even for relatively simple structures such as the beta-hairpin.



\section*{Acknowledgements}
S.S.P. gratefully acknowledges support from the Natural Sciences and
Engineering Research Council, and the A. P. Sloan Foundation. 

\bibliographystyle{plain}



\section*{Tables}

\begin{table} [hp]
\caption{${\cal D}/N$ (in units of link length squared) between the aligned
  structures in figure~\ref{figalignments}. Each of the 4 columns represents the
  structural pairs for the cost function labelled. For example,
  column~3 gives $\Dist/N$ for structural pairs in
  figure~\ref{figalignments} aligned using MRSD. }
  \centering
  \begin{tabular}[ht]{c | c | c | c| c}
& \multicolumn{4}{c}{Alignment cost function} \\
    $N$ & $\cal D$ & ${\cal D}_1$&    MRSD &   RMSD  \\ \hline
    4   & 0.785   & 0.785     & 0.785  & 0.822   \\ 
    6   & 1.391   & 1.415     & 1.473  & 1.419   \\
    8   & 1.974   & 1.983     & 2.085  & 2.014   \\
    10  & 2.559   & 2.574     & 2.654  & 2.615   \\
    12  & 3.127   & 3.158     & 3.197  & 3.216   \\
    14  & 3.674   & 3.705     & 3.726  & 3.817   \\
    16  & 4.207   & 4.235     & 4.247  & 4.418   \\
    18  & 4.732   & 4.769     & 4.762  & 5.019   \\
    20  & 5.252   & 5.294     & 5.272  & 5.620   \\
    22  & 5.767   & 5.802     & 5.783  & 6.221   \\
  \end{tabular}
  
  \label{tab:results}
\end{table}

\begin{table} [hp]
  \caption{MRSD (in units of link length) between the aligned
    structures in figure~\ref{figalignments} using the four cost
    functions we considered. For example,
  column~1 gives MRSD for structural pairs in
  figure~\ref{figalignments} aligned using the distance $\Dist$. } 
  \centering
  \begin{tabular}[ht]{c | c | c | c| c}
& \multicolumn{4}{c}{Alignment cost function} \\
    $N$ & $\cal D$ & ${\cal D}_1$&    MRSD &   RMSD  \\ \hline
    4   & 0.707    & 0.707     & 0.707   & 0.809    \\ 
    6   & 1.375    & 1.393     & 1.337   & 1.412    \\
    8   & 1.961    & 1.960     & 1.899   & 2.008    \\
    10  & 2.547    & 2.545     & 2.436   & 2.610    \\
    12  & 3.062    & 3.108     & 2.959   & 3.211    \\
    14  & 3.575    & 3.675     & 3.475   & 3.813    \\
    16  & 4.081    & 4.004     & 3.987   & 4.414    \\
    18  & 4.585    & 4.506     & 4.495   & 5.015    \\
    20  & 5.088    & 5.008     & 5.002   & 5.616    \\
    22  & 5.591    & 5.511     & 5.508   & 6.218    \\
  \end{tabular}
  \label{tab:resultsmrsd}
\end{table}

\clearpage
\section*{Figures}

\begin{figure} [hp]
  \centering
  \caption{Order parameters do not always correlate with kinetic
    proximity. Structure $A$ above is more native-like according to
    the fraction of native contacts, while structure $B$ is more
    native-like according to $RMSD$, and is also closer kinetically to
    the native structure.}
  \label{fig:QFlaw1}
\end{figure}

\end{document}